\documentclass{article}
\usepackage{times}
\usepackage{amsfonts}
\begin{document}
\noindent
{\Large ON THE CONTACT GEOMETRY AND THE POISSON GEOMETRY OF THE IDEAL GAS}
\vskip1cm
\noindent
{\bf P. Fern\'andez de C\'ordoba}${}^{a}$ and {\bf J.M. Isidro}${}^{b}$\\
Instituto Universitario de Matem\'atica Pura y Aplicada,\\ Universidad Polit\'ecnica de Valencia, Valencia 46022, Spain\\
${}^{a}${\tt pfernandez@mat.upv.es}, ${}^{b}${\tt joissan@mat.upv.es} \\
\vskip.5cm
\noindent
{\bf Abstract} We elaborate on existing notions of contact geometry and Poisson geometry as applied to the classical ideal gas. Specifically we observe that it is possible to describe its dynamics using a 3--dimensional contact submanifold of the standard 5--dimensional contact manifold used in the literature. This reflects the fact that the internal energy of the ideal gas depends exclusively on its temperature. We also present a Poisson algebra of thermodynamic operators for a quantum--like description of the classical ideal gas. The central element of this Poisson algebra is proportional to Boltzmann's constant. A Hilbert space of states is identified and a system of wave equations governing the wavefunction is found. Expectation values for the operators representing pressure, volume and temperature are found to satisfy the classical equations of state.

\section{Introduction}\label{einfuehrung}

The link between differential geometry and thermodynamics has provided deep insights. Motivated by the theory of relativity, early treatises presenting a geometric approach to thermodynamics appeared already in the 1930's \cite{TOLMAN}. More recently we have witnessed the use of Riemannian geometry \cite{MRUGALA0, MRUGALA1, QUEVEDO1, QUEVEDO2, RUPPEINER0, RUPPEINER2},  contact geometry \cite{BRAVETTI1, MRUGALA2, MRUGALA3, RAJEEV1}, Poisson and symplectic formulations \cite{RAJEEV2}, Finsler geometry \cite{MRUGALA1}, symmetric spaces \cite{BRAVETTI4} and generalised complex geometry \cite{NOI}, among others. 

Conversely the theory of gravity, a paradigmatic example of a physical theory drawing heavily on differential geometry, has achieved remarkable breakthroughs recently thanks to its extensive use of a thermodynamic approach \cite{PADDY1, PADDY2, PADDY3}; for topical reviews and more extensive references see, {\it e.g.}\/, \cite{PHILO, MOUSTOS}. 

In this paper we touch upon issues related to the contact geometry and to the Poisson geometry of the classical ideal gas. Concerning contact geometry \cite{ARNOLD}, we make the observation that its standard 5--dimensional contact manifold ${\cal M}$ can be reduced to a 3--dimensional contact submanifold. Concerning Poisson geometry \cite{ARNOLD}, we propose that {\it classical}\/ thermodynamic variables be regarded as operators on a Hilbert space of {\it quantum--like}\/ states.

{\it i)} The contact geometry of the classical ideal gas is usually described using a 5--dimensional contact manifold ${\cal M}$ that can be endowed with the local coordinates $U$ (internal energy), $S$ (entropy), $V$ (volume), $T$ (temperature) and $p$ (pressure). This description corresponds to a choice of the fundamental equation, in the energy representation, in which $U$ depends on the two extensive variables $S$ and $V$, {\it i.e.}\/, $U=U(S,V)$, the conjugate variables being
\begin{equation}
T=\frac{\partial U}{\partial S}, \qquad -p=\frac{\partial U}{\partial V}.
\label{emptp}
\end{equation}
Then the standard contact form on ${\cal M}$ reads
\begin{equation}
\alpha={\rm d}U+T{\rm d}S-p{\rm d}V.
\label{cinque}
\end{equation}
It is however well known that the internal energy of the ideal gas can be taken to depend only on the temperature $T$. This is achievable by means of 
a Legendre transformation, but it does not account for the reduction in the number of variables that $U$ depends on. In this paper we show that, in the particular case of the ideal gas, a change of variables can be identified in configuration space (coordinatised by $S$, $V$) that reduces the fundamental equation to an expression $U=U(x)$, where $x=x(S,V)$ is a single coordinate. Supplemented with its corresponding conjugate variable $p_x$, the contact form on this 3--dimensional contact submanifold ${\cal S}$ (coordinatised by $x$, $p_x$, $U$) now reads
\begin{equation}
\beta={\rm d}U+p_x{\rm d}x.
\label{extrsdici}
\end{equation}
 
{\it ii)} Concerning the Poisson geometry of the ideal gas, we propose that {\it classical}\/ thermodynamic variables be regarded as operators on a Hilbert space of {\it quantum--like}\/ states. Promoting functions to operators is in fact a natural thing to do, once fluctuations have been described using path integrals \cite{RUPPEINER1} in the spirit ref.  \cite{ONSAGER}. We call our analysis {\it quantum--like}\/, by which we mean that {\it a quantum formalism is being used in order to analyse classical thermostatics}\/. Indeed we will see that Planck's quantum of action $\hbar$ is absent altogether, its role being played instead by Boltzmann's constant $k_B$.

\section{The PDE's of state of the ideal gas}\label{opertherm}

Let us consider a number $N$ of particles of ideal gas (monoatomic for simplicity). Its fundamental equation in the energy representation $U=U(S,V)$ reads \cite{CALLEN}
\begin{equation}
U(S,V)=U_0\exp\left(\frac{2S}{3Nk_B}\right)\left(\frac{V_0}{V}\right)^{2/3},
\label{raewe}
\end{equation}
with $U_0,V_0$ certain fiducial values. Each one of the two equations defining the conjugate variables in (\ref{emptp}) qualifies as an equation of state. One can introduce Poisson brackets on the 4--dimensional Poisson manifold ${\cal P}$  (a submanifold of ${\cal M}$) spanned by the coordinates $S$, $V$ and their conjugate variables $T$, $-p$, the nonvanishing brackets being 
\begin{equation}
\{S,T\}=1, \qquad \{V,-p\}=1.
\label{fundbra}
\end{equation}
(Our definition of the Poisson brackets $\{\cdot\,,\cdot\}$ follows that of ref. \cite{ARNOLD}). Given now an equation of state 
\begin{equation}
f(p,T,\ldots)=0,
\label{aeternum}
\end{equation}
we will substitute the canonical variables (\ref{emptp}) into it, in order to obtain
\begin{equation}
f\left(-\frac{\partial U}{\partial V},\frac{\partial U}{\partial S},\ldots\right)=0.
\label{coelum}
\end{equation}
We call (\ref{coelum}) {\it a partial differential equation of state}\/ (for short, {\it PDE of state}\/). 

There are two independent conjugate variables, hence two independent equations of state. The definition of the pressure $p=-\partial U/\partial V$, plus Eq. (\ref{coranta}) below, yield the well--known law
\begin{equation}
pV=Nk_BT,
\label{corantbis}
\end{equation}
while the relation $T=\partial U/\partial S$ yields the equipartition theorem, 
\begin{equation}
U=\frac{3}{2}Nk_BT.
\label{coranta}
\end{equation}
In turn, Eqs. (\ref{corantbis}) and (\ref{emptp}) yield the first  PDE of state,
\begin{equation}
V\frac{\partial U}{\partial V}+Nk_B\frac{\partial U}{\partial S}=0,
\label{treinta}
\end{equation}
while Eqs. (\ref{coranta}) and (\ref{emptp}) yield the second  PDE of state
\begin{equation}
U-\frac{3}{2} Nk_B\frac{\partial U}{\partial S}=0.
\label{veintiuno}
\end{equation}
One readily integrates the system (\ref{treinta}) and (\ref{veintiuno}) to obtain the fundamental equation (\ref{raewe}) we started off with.

The successive changes of variables
\begin{equation}
v:=\ln \left(\frac{V}{V_0}\right), \quad s:=\frac{S}{Nk_B}
\label{kange}
\end{equation}
and
\begin{equation}
x:=s-v,\qquad y:=s+v,
\label{bllko}
\end{equation}
transform the PDE's of state (\ref{treinta}) and (\ref{veintiuno}) into 
\begin{equation}
\frac{\partial U}{\partial y}=0
\label{difel}
\end{equation}
and
\begin{equation}
U-\frac{3}{2}\frac{\partial U}{\partial x}=0
\label{litto}
\end{equation}
respectively. The solution to (\ref{difel}) and (\ref{litto}) reads
\begin{equation}
U=U(x)=U_0\exp\left(\frac{2x}{3}\right).
\label{koppla}
\end{equation}
In particular, $U$ does not depend on $y$. This reflects the well--known fact that the internal energy of an ideal gas can be taken to depend exclusively on the temperature $T$. It must be realised, though, that the change to $T$ as {\it the one}\/ independent variable involves a Legendre transformation in phase space, which is not the case in (\ref{kange}), (\ref{bllko}). Instead, the above change of variables can be performed without exiting configuration space.  Computing the conjugate variable $p_x$ corresponding to the variable $x$ we find
\begin{equation}
p_x=\frac{\partial U}{\partial x}=\frac{2}{3}U.
\label{pequis}
\end{equation}
Since $x$ is dimensionless, $p_x$ has the dimensions of energy. Now comparing to the temperature $T$ as computed from Eq. (\ref{raewe}),
\begin{equation}
T=\frac{\partial U}{\partial S}=\frac{2}{3Nk_B}U,
\label{teperat}
\end{equation}
we arrive at the equality of the two variables $p_x$ and $T$, modulo Boltzmann's constant to correct the different dimensionalities:
\begin{equation}
p_x=Nk_BT.
\label{paques}
\end{equation}
The other canonical variable, $p_y=\partial U/\partial y$, vanishes identically by virtue of the PDE of state (\ref{difel}). 

We are thus left with a 3--dimensional contact submanifold ${\cal S}$ (coordinatised by $x$, $p_x$ and $U$) of the initial 5--dimensional contact manifold ${\cal M}$ (coordinatised by $U$, $S$, $V$, $T$ and $-p$). The contact form $\beta$ on ${\cal S}$ given in Eq. (\ref{extrsdici}) can be readily shown to equal the restriction, to the submanifold ${\cal S}$, of the contact form $\alpha$ on ${\cal M}$ given in Eq. (\ref{cinque}):
\begin{equation}
\alpha=\beta\; {\rm on}\; {\cal S}.
\label{kontakt}
\end{equation}

\section{The wave equations of the ideal gas}\label{amsams}

Application of the canonical quantisation rules to the PDE (\ref{coelum}) will produce a quantum--like wave equation. Initially, this canonical quantisation will be carried out on a 2--dimensional Lagrangian submanifold (the configuration space spanned by $S$ and $V$) of the 5--dimensional contact manifold ${\cal M}$. At a later stage will we apply the changes of variables (\ref{kange}) and (\ref{bllko}) in order to directly quantise a 1--dimensional Lagrangian submanifold (the configuration space spanned by $x$) of the 3--dimensional contact manifold ${\cal S}$. The two procedures (quantisation and reduction) will be seen to commute.

\subsection{Quantum commutators}\label{bhvfeury}

Promoting the real functions $S$, $V$, $T$, $p$ to Hermitian operators on a Hilbert space, the Poisson algebra of functions (\ref{fundbra}) becomes the Poisson--Lie algebra of operator commutators
\begin{equation}
[S, T]=q{\bf 1},\qquad [V,- p]=q{\bf 1}, \qquad q\in\mathbb{C},
\label{brafund}
\end{equation}
where $[X,Y]:=XY-YX$ by definition. The right--hand side contains a quantum of energy $q$ and the identity operator ${\bf 1}$. For reasons that will soon become apparent we will assume our ideal gas to be in thermal equilibrium with a bath kept at a constant temperature $T_B$, so we can write
\begin{equation}
q=zNk_BT_B,
\label{refete}
\end{equation}
where $k_B$ is Boltzmann's constant and $z\in\mathbb{C}$ a free parameter. 

Assume that $S$ and $V$ are represented as multiplication operators on quantum--like wavefunctions $\psi(S,V)$. Then the conjugate variables act on $\psi$ by differentiation, 
\begin{equation}
T=-q\frac{\partial}{\partial S}, \qquad p=q\frac{\partial}{\partial V}.
\label{merto}
\end{equation}
Modulo operator--ordering ambiguities, substituting the variables (\ref{merto}) into the equations of state (\ref{corantbis}) and (\ref{coranta}) produces two time--independent wave equations that quantum--like states $\psi(S,V)$ must satisfy. As we will prove, the solutions to these wave equations will automatically be $L^2$--integrable on configuration space 
$[S_0,S_1]\times[V_0,V_1]$. Square integrability of $\psi$ ensures that $\vert\psi(S,V)\vert^2$ belongs to $L^1([S_0,S_1]\times[V_0,V_1])$, as befits a thermodynamic probability distribution.

\subsection{Thermodynamic wave equations}

Next we apply the quantum rules (\ref{merto}) to the equations of state (\ref{corantbis}) and (\ref{coranta}). Three possible orderings for the operators $V$ and $p$ are $Vp$, $pV$ and Weyl's $(Vp+pV)/2$; the simplest one, that we will use here, is $Vp$. Therefore quantum--like states will be simultaneous solutions to 
\begin{equation}
\left(V\frac{\partial}{\partial V}+Nk_B\frac{\partial}{\partial S}\right)\psi(S,V)=0
\label{bernie}
\end{equation}
and
\begin{equation}
\left(U(S,V)+\frac{3q}{2}Nk_B\frac{\partial}{\partial S}\right)\psi(S,V)=0.
\label{veinticuatro}
\end{equation}
The above system is solved by
\begin{equation}
\psi_q(S,V)=\exp\left(-\frac{1}{q}U(S,V)\right)
\label{veinticinco}
\end{equation}
with an arbitrary value of the parameter $q=zNk_BT_B$; equivalentely $z$ runs over the whole complex plane $\mathbb{C}$.

Some particular choices of $z$ in (\ref{veinticinco}) deserve attention. The value $z=1$ leads to the real exponential
\begin{equation}
\psi_{z=1}(S,V):=\exp\left(-\frac{U(S,V)}{Nk_BT_B}\right),
\label{cincuentacuatrobis}
\end{equation}
while $z={\rm i}$ yields the complex exponential
\begin{equation}
\psi_{z={\rm i}}(S,V):=\exp\left({\rm i}\frac{U(S,V)}{Nk_BT_B}\right).
\label{cincuentacuatro}
\end{equation}
The above states are distinguished in the sense that they are reminiscent of analogous exponentials arising in statistical mechanics and quantum mechanics, respectively.

\subsection{The wave equations on the reduced phase space}

At the quantum--like level, the reduction of the phase space from the 5--dimensional ${\cal M}$ to the 3--dimensional ${\cal S}$ takes place exactly as at the classical level. Namely, the change of variables (\ref{kange}), (\ref{bllko}) transforms the system (\ref{bernie}), (\ref{veinticuatro}) into
\begin{equation}
\frac{\partial\psi}{\partial y}=0
\label{sanderz}
\end{equation}
and
\begin{equation}
\left(U(x)+\frac{3q}{2}\frac{\partial}{\partial x}\right)\psi=0,
\label{bllka}
\end{equation}
where $U(x)$ was already identified in Eq. (\ref{koppla}). The above is solved by 
\begin{equation}
\psi_q(x)=\exp\left(-\frac{1}{q}U(x)\right),
\label{hhdfr}
\end{equation}
in agreement with our previous result (\ref{veinticinco}). In particular, the $y$--dependence drops out as before. Now the internal energy is defined modulo the addition of a real constant,
\begin{equation}
U(x)\rightarrow U(x)+C, \qquad C\in\mathbb{R}.
\label{modlo}
\end{equation}
This symmetry reflects the global invariance of the (unnormalised) wavefunction $\psi_q$ under the action of $\mathbb{C}^*$, the multiplicative group of nonzero complex numbers:
\begin{equation}
\psi_q(x)\rightarrow\exp(-C/q)\psi_q(x).
\label{imba}
\end{equation}
As announced above the two procedures, {\it quantisation}\/ and {\it reduction}\/, commute in our case.

\section{Discussion}\label{diskku}

Classical thermostatics is governed by a system of  partial differential equations of state, the integral of which yields the fundamental equation of the thermodynamic system under consideration. In the case of the ideal gas, the PDE's of state are Eqs.  (\ref{treinta}) and (\ref{veintiuno}), their solution being given by Eq. (\ref{raewe}). 

In this work we have established that the classical PDE's of state for the ideal gas have a set of quantum--like counterparts, the wave equations (\ref{bernie}) and (\ref{veinticuatro}). Their integral, Eq. (\ref{veinticinco}), is the wavefunction of a quantum--like description of the ideal gas. Summarising one can say that {\it the quantum--like wavefunction is the exponential of the classical fundamental equation of the ideal gas}\/. This is in perfect analogy with corresponding notions in WKB quantum mechanics. 

One can find a coordinate change that transforms the relevant equations into a reduced form. Under {\it reduced}\/ we understand that the equations {\it involve the least number of variables}\/; geometrically this corresponds to the reduction from the initial 5--dimensional phase space ${\cal M}$ to a 3--dimensional subspace ${\cal S}$. For the classical PDE's of state this reduced form is (\ref{litto}); its corresponding quantum--like equation is (\ref{bllka}). Instead of two real coordinates $S$, $V$ one is left with just one coordinate $x$ that the fundamental equation $U=U(x)$ and the wavefunction $\psi_q=\psi_q(x)$ depend on. We have succeeded in finding a coordinate transformation {\it without exiting configuration space}\/  (see Eqs. (\ref{kange}), (\ref{bllko})) that reduces the number of independent variables the internal energy depends on. So the phase space of the classical ideal gas is a 3--dimensional contact submanifold ${\cal S}$ of the standard 5--dimensional contact manifold ${\cal M}$. This reduction is a feature of the classical ideal gas that need not (and generally will not) hold for other thermodynamic systems. We should stress, however, that this dimensional reduction from 5 to 3 implies information loss. Inverting this dimensional reduction ({\it i.e.}\/, returning from 3 to 5 dimensions) cannot be done without prior knowledge of the equation of state.

The quantum commutators (\ref{brafund}) lead to the uncertainty relations $\Delta S\Delta T\geq \vert q\vert/2$, $\Delta p\Delta V\geq \vert q\vert/2$. As opposed to the quantum--mechanical uncertainty relation $\Delta x\Delta p\geq \hbar/2$, the quantum $q$ carries the dimensions of energy (Boltzmann's constant $k_B$ multiplied by the temperature $T_B$ of the bath). Moreover, since $T_B$ is arbitrary, the quantum $q$ may be taken to be arbitrarily small. This is a fundamental difference with respect to quantum mechanics. A model containing both Planck's constant $\hbar$ and Boltzmann's constant $k_B$ has been considered in ref. \cite{SASA}.

{}For any fixed value of the central element $q$ in the quantum Poisson algebra, the space of solutions to the wave equation is a 1--dimensional subspace of the Hilbert space $L^1([S_0,S_1]\times [V_0,V_1])$. Moreover, there is a whole $\mathbb{C}$'s worth of central elements $q$ for the quantum Poisson algebra (\ref{brafund}). In quantum theory, the Hilbert space $L^2([S_0,S_1]\times [V_0,V_1])$ provides a {\it unitary}\/ representation of the quantum Poisson algebra (\ref{brafund}); for this it is necessary (though not sufficient) that the quantum $q$ be pure imaginary \cite{THIRRING}. Unitarity of this representation implies that observable quantities are represented by Hermitian operators. Thus unitarity is ruled out for the quantum states  with $q\in\mathbb{R}$, such as the state (\ref{cincuentacuatrobis}). How does the the state (\ref{cincuentacuatro}) fare?

Let us recall \cite{THIRRING} that periodic boundary conditions on the wavefunction,  $\psi(a)=\psi(b)$, ensure hermiticity of $-{\rm i}\partial/\partial x$ on $L^2([a,b])$. In the absence of periodicity, the Hermitian property is generally not guaranteed. However the state (\ref{cincuentacuatro}) is not periodic on $[S_0,S_1]\times[V_0,V_1]$. The inevitable conclusion is that, whatever the value of $q$, we are forced to deal with a non--Hermitian theory. Now the Hermitian property of operators is a sufficient (but not a necessary) condition to ensure that expectation values are real \cite{THIRRING}. Fortunately, the relevant expectation values in the states (\ref{veinticinco}) are all real. This follows by taking the scalar product of Eqs. (\ref{bernie}) and (\ref{veinticuatro}) with $\psi_q$. In this way classicality is recovered in the form of Eherenfest's theorem \cite{LANDAU3} for the expectation values of the relevant operators entering the equations of state.

In the presence of a gravitational field, thermal fluctuations have been argued to be indistinguishable from quantum fluctuations \cite{PADDY0}. 
Admittedly, {\it quantum}\/ is Hermitean while {\it thermal}\/ is not, and the loss of unitarity reported above reflects this fact. In this work we have shown that, in the thermodynamics of the ideal monoatomic gas, one can go a long way replacing {\it thermal}\/ with {\it quantum}\/ without noticing the difference.

\vskip.5cm
\noindent
{\bf Acknowledgements}  This research was supported by grant no. ENE2015-71333-R (Spain).


\begin{thebibliography}{99}

\bibitem{ARNOLD}
V. Arnold, {\it Mathematical Methods of Classical Mechanics}, Graduate Texts in Mathematics {\bf 60}, Springer, Berlin (1989). 

\bibitem{BRAVETTI1}
A. Bravetti, {\it Contact Hamiltonian Dynamics: the Concept and its Use}, Entropy {\bf 19} (2017) 535.

\bibitem{BRAVETTI4}
A. Bravetti, C. L\'opez-Monsalvo and H. Quevedo, {\it Maximally Symmetric Spacetimes Emerging from Thermodynamic Fluctuations}, {\tt arXiv:1503.08358 [gr-qc]}.

\bibitem{CALLEN}
H. Callen, {\it Thermodynamics}, Wiley, New York (1960).

\bibitem{NOI}
P. Fern\'andez de C\'ordoba and J.M. Isidro, {\it Generalised Complex Geometry in thermodynamic Fluctuation Theory}, Entropy {\bf 17} (2015) 5888, {\tt arXiv:1505.06647 [math-ph]}.

\bibitem{PADDY0}
S. Kolekar and T. Padmanabhan, {\it Indistinguishability of Thermal and Quantum Fluctuations}, Class. Quant. Grav. {\bf 32} (2015) 202001, {\tt arXiv:1308.6289 [gr-qc]}. 

\bibitem{LANDAU3}
L. Landau and E. Lifshitz, {\it Quantum Mechanics}, vol. 3 of {\it Course of Theoretical Physics}, Butterworth--Heinemann, Oxford (2000).

\bibitem{PHILO}
N. Linnemann and M. Visser, {\it Hints towards the Emergent Nature of Gravity}, {\tt arXiv:1711.10503 [physics.hist-ph]}.

\bibitem{MOUSTOS}
D. Moustos, {\it Gravity as a Thermodynamic Phenomenon}, {\tt arXiv:1701.08967 [gr-qc]}.

\bibitem{MRUGALA0}
R. Mrugala, {\it Geometric Formulation of Equilibrium Phenomenological Thermodynamics},  Rep. Math. Phys. {\bf 14} (1978) 419.

\bibitem{MRUGALA1}
R. Mrugala, {\it Riemannian and Finslerian Geometry in Thermodynamics}, Open Sys. \& Inf. Dyn. {\bf 1} (1992) 379. 

\bibitem{MRUGALA2}
R. Mrugala, J. Nulton, J. Sch\"on and P. Salamon, {\it Statistical Approach to the Geometric Structure of Thermodynamics}, Phys. Rev. {\bf A41} (1990) 3156.  

\bibitem{MRUGALA3}
R. Mrugala, J. Nulton, J. Sch\"on and P. Salamon, {\it Contact Structure in Thermodynamic Theory}, Rep. Math. Phys. {\bf 29} (1991) 109.

\bibitem{ONSAGER}
L. Onsager and S. Machlup, {\it Fluctuations and Irreversible Processes}, Phys. Rev. {\bf 91} (1953) 1505.

\bibitem{PADDY1}
T. Padmanabhan, {\it Thermodynamic Aspects of Gravity: New Insights}, Rept. Prog. Phys. {\bf 73} (2010) 046901, {\tt arXiv:0911.5004 [gr-qc]}.

\bibitem{PADDY2}
T. Padmanabhan, {\it General Relativity from a Thermodynamic Perspective}, Gen. Rel. Grav. {\bf 46} (2014) 1673, {\tt arXiv:1312.3253 [gr-qc]}.

\bibitem{PADDY3}
T. Padmanabhan, {\it Gravity and/is Thermodynamics},  Curr. Sci. {\bf 109} (2015) 2236, {\tt arXiv:1512.06546 [gr-qc]}.

\bibitem{QUEVEDO1}
H. Quevedo and A. V\'azquez, {\it The Geometry of Thermodynamics},  AIP Conf. Proc. {\bf 977} (2008) 165, {\tt arXiv:0712.0868 [math-ph]}.

\bibitem{QUEVEDO2}
H. Quevedo, A. S\'anchez, S. Taj and A. V\'azquez, {\it Curvature as a Measure of the Thermodynamic Interaction}, {\tt arXiv:1011.0122 [gr-qc]}.

\bibitem{RAJEEV1}
G. Rajeev, {\it Quantization of Contact Manifolds and Thermodynamics}, Ann. Phys. {\bf 323} (2008) 768,  {\tt arXiv:math-ph/0703061}.

\bibitem{RAJEEV2}
G. Rajeev, {\it A Hamilton--Jacobi Formalism for Thermodynamics}, Ann. Phys. {\bf 323} (2008) 2265, {\tt arXiv:0711.4319 [hep-th]}.

\bibitem{RUPPEINER0}
G. Ruppeiner, {\it Thermodynamics: a Riemannian Geometric Model}, Phys. Rev. {\bf A20} (1979) 1608.

\bibitem{RUPPEINER1}
G. Ruppeiner, {\it New Thermodynamic Fluctuation Theory Using Path Integrals}, Phys. Rev. A {\bf 27} (1983) 1116.

\bibitem{RUPPEINER2}
G. Ruppeiner, {\it Riemannian Geometry in Thermodynamic Fluctuation Theory}, Rev. Mod. Phys. {\bf 67} (1995) 605.

\bibitem{SASA}
S. Sasa and Y. Yokokura, {\it Thermodynamic Entropy as a Noether Invariant}, Phys. Rev. Lett. {\bf 116} (2016) 140601, {\tt arXiv:1509.08943 [cond-mat]}.

\bibitem{THIRRING}
W. Thirring, {\it Quantum Mathematical Physics}, Springer, Berlin (2003).

\bibitem{TOLMAN}
R. Tolman, {\it Relativity, Thermodynamics and Cosmology}, Dover, New York (1987).




\end{thebibliography}
\end{document}